\documentclass[12pt,a4paper,fleqn]{article}
\usepackage[DIV12]{typearea}
\usepackage{amsmath,amsfonts,amssymb}
\usepackage{graphicx}
\usepackage{hyperref}
\usepackage{cite}
\usepackage{mathrsfs}
\usepackage{bbm}

\newcommand{\eVdist}{\kern-0.06em}

\DeclareMathOperator{\diag}{diag}

\newcommand{\SU}[1]{\ensuremath{\mathrm{SU}(#1)}}
\newcommand{\U}[1]{\ensuremath{\mathrm{U}(#1)}}
\newcommand{\Z}[1]{\ensuremath{\mathbbm{Z}_{#1}}} 

\numberwithin{equation}{section}
\numberwithin{table}{section}

\allowdisplaybreaks[1]

\title{The Hilbert basis method for $\boldsymbol{D}$--flat directions and the
superpotential}

\begin{document}

\begin{titlepage}
\vspace*{-1cm}
\begin{flushright}
TUM-HEP 814/11\\
MPP-2011-99\\
CERN-PH-TH/2011-193
\end{flushright}

\vspace*{0.5cm}

\begin{center}
{\Large\bf
The Hilbert basis method for $\boldsymbol{D}$--flat directions\\[0.2cm] and the
superpotential
}

\vspace{0.7cm}

\textbf{
Rolf Kappl
$^{\,a,b}${},
Michael Ratz
$^{\,a,c}${},
Christian Staudt
$^{\,a}${}
}
\\[8mm]
\textit{$^a$\small
~Physik-Department T30, Technische Universit\"at M\"unchen, \\
James-Franck-Stra\ss e, 85748 Garching, Germany
}
\\[5mm]
\textit{$^b$\small
~Max-Planck-Institut f\"ur Physik (Werner-Heisenberg-Institut),\\
F\"ohringer Ring 6,
80805 M\"unchen, Germany
}
\\[5mm]
\textit{$^c$\small
~Theory Group, CERN, 1211 Geneva 23, Switzerland
}
\end{center}

\vspace{0.4cm}

\begin{abstract}
We discuss, using the Hilbert basis method, how to efficiently construct a
complete basis for $D$-flat directions in supersymmetric Abelian and
non--Abelian gauge theories. We extend the method to discrete ($R$ and non--$R$)
symmetries. This facilitates the construction of a basis of all superpotential
terms in a theory with given symmetries. 
\end{abstract}

\end{titlepage}

\newpage

\section{Introduction}

Holomorphic gauge invariant monomials play an important role in the understanding of
supersymmetric theories and phenomenological applications. They represent
$D$--flat directions in supersymmetric gauge theories \cite{Buccella:1982nx} and
constitute possible superpotential terms. However, in somewhat complex theories
the explicit constructions of these monomials can be quite cumbersome in
practice. For instance, already the construction of all gauge invariant
monomials for the minimal supersymmetric extension of the standard model (MSSM)
is rather involved \cite{Gherghetta:1995dv} (see also \cite{Cleaver:1997jb} for
the discussion in stringy extensions of the MSSM). 

In \cite{Kappl:2010yu} it was shown how to construct the so--called Hilbert
basis for holomorphic gauge invariant monomials $\mathscr{M}$, which are known
to be in one--to--one correspondence with the $D$--flat directions. The Hilbert
basis allows us to write every monomial $\mathscr{M}$ in the  form
\begin{equation}
\label{eq:hilintro}
 \mathscr{M}~=~\prod\limits_{i=1}^H \mathscr{M}_i^{\eta_i}
 \qquad\text{with}~\eta_i~\in~\mathbbm{N}_0\;.
\end{equation}
Here $H$ is the number of independent basis monomials $\mathscr{M}_i$ which
can only be determined algorithmically.

The purpose of this note is to extend the notion of the Hilbert basis such as to
include discrete $R$ and non--$R$ symmetries as well. In
section~\ref{sec:ReviewHilbert} we start by reviewing the Hilbert basis method
for continuous gauge symmetries. Section~\ref{sec:DiscreteSymmetries} is devoted
to the extension to discrete symmetries. The general case is discussed in
section~\ref{sec:AllTogether}. In section~\ref{sec:Applications} we comment on
potential applications, and finally, section~\ref{sec:Conclusions} contains our
conclusions.

\section{Review of Hilbert bases for continuous gauge symmetries}
\label{sec:ReviewHilbert}

Let us briefly review the Hilbert basis method for (continuous) gauge symmetries
\cite{Kappl:2010yu}. We start by looking at a theory with a single \U1 factor
and then extend the discussion to $L$ different $\U1$ factors.

\subsection{Warm--up example: a single U(1)}

Consider a \U1 gauge theory and fields $\phi^{(f)}$ ($1\le f\le F$) with charges
$q^{(f)}$. A monomial
\begin{equation}\label{eq:Monomial}
 \mathscr{M}~=~\left(\phi^{(1)}\right)^{n_1}\cdots\left(\phi^{(F)}\right)^{n_F}
\end{equation}
is gauge invariant if 
\begin{subequations}\label{eq:GaugeInvarianceU1}
\begin{equation}
q^{(1)}\, n_1+\dots+q^{(F)}\, n_F ~=~0\;.
\end{equation}
This condition may be recast as
\begin{equation}
 q^{T}\cdot n~=~0\;,
\end{equation}
\end{subequations}
$q^T=(q^{(1)},\dots,q^{(F)})$ and  $n^T=(n_1,\dots,n_F)$. That is, the vector
$n$ has to be orthogonal to the charge vector $q$.  The requirement that
$\mathscr{M}$ be holomorphic amounts to demanding that $n_i\in\mathbbm{N}_0$. 
The solutions are the intersection of the hyperplane perpendicular to $q$ and
the lattice points in $\mathbbm{N}_0^F$. The so--called Hilbert basis 
\begin{equation}
 \mathscr{H}~=~\left\{h^{(1)},\dots h^{(H)}\right\},
\end{equation}
is a complete set of vectors $h^{(i)}$ with the property that each solution $n$
of \eqref{eq:GaugeInvarianceU1} can be written as
\begin{equation}
 n~=~\sum\limits_{i=1}^H\eta_i\,h^{(i)}
 \qquad\text{with}~\eta_i~\in~\mathbbm{N}_0\;.
\end{equation}
Every element $h^{(i)}$ of the Hilbert basis is in one--to--one 
correspondence with a gauge invariant monomial 
\begin{equation}
\mathscr{M}_i~=~\left(\phi^{(1)}\right)^{h^{(i)}_1}\cdots
\left(\phi^{(F)}\right)^{h^{(i)}_F}
\end{equation}
such that every gauge invariant monomial is given by 
\eqref{eq:hilintro}.

\subsection{Generalization to $\boldsymbol{L}$ U(1) factors}

In the case of $L$ \U1 factors the condition \eqref{eq:GaugeInvarianceU1} can be
rewritten as
\begin{equation}
 Q\cdot n~=~0,\quad n\in\mathbbm{N}_0^F
\end{equation}
with the charge matrix
\begin{equation}\label{eq:MatrixQ}
 Q~=~\left(\begin{array}{ccc}
  q_1^{(1)} & \cdots & q_1^{(F)}\\
  \vdots & & \vdots\\
  q_L^{(1)} & \cdots & q_L^{(F)}
 \end{array}\right)\;,
\end{equation}
where $q_\ell^{(f)}$ denotes the $\ell^\mathrm{th}$ \U1 charge of the
$f^\mathrm{th}$ field $\phi^{(f)}$.

An important fact about the above problem is that it is well--known in the
mathematical literature \cite{Schrijver86,hemmecke:2002,bruns:2009}.  There are
efficient algorithms such as the ones provided by \cite{4ti2,Normaliz}, allowing
us to compute the Hilbert basis for a given matrix $Q$ very efficiently.

\subsection{Non--Abelian symmetries}
\label{sec:NonAbelian}

In the case of a non--Abelian symmetry $G$, gauge invariance of a monomial
composed of $G$ representations $\boldsymbol{r}^{(i)}$ is equivalent to gauge
invariance w.r.t.\ the $r$ \U1 factors generated by the Cartan generators of
$G$. That is, the Hilbert basis method allows us immediately to construct $G$
invariant monomials. Some of the resulting monomials are zero and others are
redundant. Examples for vanishing monomials include the baryons of \SU{N_c}
theories, $\varepsilon^{\alpha_1\alpha_2\cdots\alpha_{N_c}}\,  
\phi^{(i_1)}_{\alpha_1}\,\phi^{(i_2)}_{\alpha_2}\cdots
\phi^{(i_{N_c})}_{\alpha_{N_c}}$, which vanish if, say, $i_1=i_2$. In order to
construct only non--vanishing and inequivalent monomials, there are various
methods available. One of them is `consecutive basis building' and the other is
to systematically remove redundant monomials from the outcome of the Hilbert
basis construction. Both methods are briefly reviewed in
appendix~\ref{app:NonAbelianMonomials}. A  \texttt{mathematica} package allowing
for an automatized computation of the independent monomials can be found at
\cite{Dcode:2011df}.

\section{Discrete symmetries}
\label{sec:DiscreteSymmetries}

In what follows, we generalize the Hilbert basis method such as to applicable to
discrete symmetries also. This allows us, in particular, to identify a complete
basis for allowed superpotential terms. 

\subsection{Discrete non--$\boldsymbol{R}$ symmetries}

We now discuss how to construct the Hilbert basis for discrete non--$R$
symmetries. We illustrate our method by a simple example, which we then
generalize.

\subsubsection{Warm--up example}

We start by a single $\Z{M}$ symmetry under which the $\phi^{(f)}$ have charges
$p^{(f)}$. The requirement of $\Z{M}$ invariance of the monomial
\eqref{eq:Monomial} translates into
\begin{equation}\label{eq:ZM1}
 p^{(1)}\,n_1+\dots+p^{(F)}\,n_F~=~0\mod M\;.
\end{equation}
Without loss of generality we can assume that all discrete charges $p^{(f)}$ are
non--negative. Equation~\eqref{eq:ZM1}  is equivalent to 
\begin{equation}
 \left(-M,p^{(1)},\dots p^{(F)}\right)
 \cdot
 \left(\begin{array}{c}
  m\\
  n_1\\
  \vdots\\
  n_F
 \end{array}\right)~=~0
 \qquad\text{with}~m~\in~
 \mathbbm{N}_0
\end{equation}
and $n_f\in\mathbbm{N}_0$, as before.  One can now compute the Hilbert basis for
the above problem. The basis monomials are then given by
\begin{equation}
\mathscr{M}_i~=~\left(\phi^{(1)}\right)^{h^{(i)}_2}\cdots
\left(\phi^{(F)}\right)^{h^{(i)}_{F+1}}
\end{equation}
after truncation of the first element $h^{(i)}_1$ which represents not a 
field $\phi^{(f)}$ but $m$.

\subsubsection{Multiple discrete non--$\boldsymbol{R}$ symmetries}

The extension to $\Z{M_1}\times\dots\times\Z{M_K}$ is straightforward.  We
assume that $\Z{M_1}\times\dots\times\Z{M_K}$ is already  the smallest
irreducible symmetry of a possible larger discrete group (cf.\
\cite{Petersen:2009ip}). We define the charge matrix
\begin{equation}
 C~=~\left(-M~|~P\right)\;,
\end{equation}
where
\begin{equation}\label{eq:MatrixM}
 M~=~\diag\left(M_1,\dots, M_K\right)
\end{equation}
and
\begin{equation}\label{eq:MatrixP}
 P~=~\left(\begin{array}{ccc}
  p_1^{(1)} & \dots & p_1^{(F)}\\
  \vdots & & \vdots \\
  p_K^{(1)} & \dots & p_K^{(F)}
 \end{array}\right)
\end{equation}
with $p_k^{(f)}$ denoting the $\Z{M_k}$ charge of $\phi^{(f)}$.  We can again
assume that all charges $p_k^{(f)}$ are non--negative.  

\subsection{Discrete $\boldsymbol{R}$ symmetries}

Let us now turn to discrete $R$ symmetries.  We start by discussing a single
$\Z{N}^R$ and then generalize the setting to more than one discrete $R$
symmetry.

\subsubsection{Warm--up example}

We start by a single $\Z{N}^R$ symmetry under which the $\phi^{(f)}$ have 
charges $r^{(f)}$, which, again, can all be chosen non--negative.  The
requirement of $\Z{N}^R$ invariance of the monomial
\eqref{eq:Monomial} translates into
\begin{equation}
 r^{(1)}\,n_1+\dots+r^{(F)}\,n_F~=~2\mod N\;.
\end{equation}
Here we have adopted the convention that the superpotential has $R$ charge 2.
The above equation is equivalent to 
\begin{equation}\label{eq:probleq1}
 \left(-2,-N,p^{(1)},\dots p^{(F)}\right)
 \cdot
 \left(\begin{array}{c}
  \ell\\
  m\\
  n_1\\
  \vdots\\
  n_F
 \end{array}\right)~=~0
 \qquad\text{with}~\ell~=~1\;,
\end{equation} 
$m\in\mathbbm{N}_0$ and $n_f\in\mathbbm{N}_0$, as before. One can now compute
the solution of the above problem. First, one identifies a basis of all vectors
orthogonal to $\left(-2,-N,p^{(1)},\dots p^{(F)}\right)^T$.  
To deal with the restriction $\ell~=~1$ we identify two subsets of the Hilbert
basis. One, where the first entry equals zero, which we call the homogeneous
solution space. And one where the first entry equals one, which we call the
inhomogeneous solution space. If a homogeneous solution is labeled by
$n_\mathrm{hom}^{(h)}\in\mathscr{H}$ and a inhomogeneous solution by 
$n_\mathrm{inhom}^{(i)}\in\mathscr{H}$ we find the general solution to
\eqref{eq:probleq1} to be
\begin{equation}\label{eq:GeneralSolution1}
 n~=~n_\mathrm{inhom}^{(i)}
 +\sum\limits_{h=1}^{H_0}\eta_h\,n_\mathrm{hom}^{(h)}
\end{equation} 
with fixed $i$ and $\eta_h\in\mathbbm{N}_0$. That is, there are $H_1$ branches
of solutions, where $H_1$ denotes the number of inhomogeneous solutions. Each
branch consists of $H_0$ solutions with $H_0$ denoting the number of homogeneous
solutions. We can find all monomials spanning the  superpotential by truncating
the vectors $n$ accordingly.

To see what that means in practice, consider a setting with a $\Z4^R$ symmetry
and fields $\psi$ and $\phi$ with $R$ charges $r_\psi=1$ and $r_\phi=0$. The
generalized charge matrix for this example reads
\begin{equation}
 C~=~\left(-2,-4,1,0\right)
 \;,
\end{equation}
where the last two entries are the $R$ charges of $\psi$ and $\phi$. There is
only one vector orthogonal to $C$ with first component equal to 1,  namely
$\left(1,0,2,0\right)^{T}$, such that the unique inhomogeneous solution is given
by  $n_\mathrm{inhom}~=~\left(1,0,2,0\right)^{T}$. Similarly,
we obtain two vectors orthogonal to $C$ with vanishing first component, namely
$\left(0,1,4,0\right)^{T}$ and $\left(0,0,0,1\right)^{T}$. The  entries of the
vectors represent the exponents of the fields in the corresponding monomials. We
hence have found that all allowed superpotential terms are of the form
\begin{equation}
 \mathscr{M}~=~\psi^2\,\psi^{4\eta_1}\,\phi^{\eta_2}
\end{equation}
with $\eta_i \in\mathbbm{N}_0$. Of course, one could have obtained the result
without the Hilbert basis method; however, for more complex systems this method
is highly advantageous.

\subsubsection{Multiple discrete $\boldsymbol{R}$ symmetries}

Let us extend the discussion to $\Z{N_1}^R\times\dots\Z{N_J}^R$.  Define the
charge matrix
\begin{equation}
C~=~
\left(\begin{array}{c|c|c}
 \begin{tabular}{c}
  -2 \\
  \vdots\\
  -2
  \end{tabular} & -N &  R \\
 \end{array}\right)\;,
\end{equation}
with 
\begin{equation}\label{eq:MatrixN}
 N~=~\diag\left(N_1,\dots ,N_J\right)
\end{equation}
and 
\begin{equation}\label{eq:MatrixR}
 R~=~
 \left(\begin{array}{ccc}
  r_1^{(1)} & \cdots & r_1^{(F)} \\
  \vdots &  & \vdots \\
  r_J^{(1)} & \cdots & r_J^{(F)}
 \end{array}\right)\;.
\end{equation}
Here $r_j^{(f)}$ denotes the $j^\mathrm{th}$ $R$ charge of the $f^\mathrm{th}$
field. We compute the kernel of the matrix $C$. 
The last $F$ components of the kernel vectors with the first entry equal to 1 or
0 (before truncation) define the inhomogeneous or homogeneous solutions,
respectively. The general solution will again be of the form
\eqref{eq:GeneralSolution1}. As before, the identification of the (last $F$)
entries of the vectors with the exponents will then give us the desired
invariant monomials, i.e.\ allowed superpotential terms.

\section{Putting all together}
\label{sec:AllTogether}

\subsection{Charge matrix for $\boldsymbol{\U1^L}$ symmetry with discrete
$\boldsymbol{R}$ and non--$\boldsymbol{R}$ symmetries}

We consider now the general
$\U1^L\times\Z{M_1}\times\dots\times\Z{M_K}\times\Z{N_1}^R\times\dots\Z{N_J}^R$
case. The charge matrix is
\begin{equation}
 C ~=~
 \left(\begin{array}{c|c|c|c}
 \begin{tabular}{c}
  -2 \\
  \vdots\\
  -2
  \end{tabular} & -N &  0 & R \\
 \hline
 0 & 0 &  -M &   P \\
 \hline
 0 & 0 & 0 &  Q 
 \end{array}\right)
\end{equation} 
with $Q$, $P$, $R$, $N$ and $M$ defined in equations \eqref{eq:MatrixQ},
\eqref{eq:MatrixP}, \eqref{eq:MatrixR}, \eqref{eq:MatrixN} and
\eqref{eq:MatrixM}, respectively.  We compute the kernel of $C$, and, again, we
decompose these vectors in those with first components equal to 1 or 0. From
those we obtain the inhomogeneous or homogeneous solutions, respectively, by
projecting on the last $F$ components. As before, the general solution is of the
form \eqref{eq:GeneralSolution1}, and identifying the entries of the vectors
with the exponents will then give us the desired invariant monomials, i.e.\
allowed superpotential terms.

\subsection{A stringy example}

We base our example on the $\Z2\times\Z2$ orbifold discussed in
\cite{Kappl:2010yu}. We consider the $\Z4^R$ vacuum discussed there and
construct the superpotential for the standard model singlets with $R$ charges 0
or 2. At the orbifold point, the symmetry seen by these fields is of the type
\begin{equation}
 G_\mathrm{symm}~=~\U1^L\times (\Z{2})^6\times(\Z{4}^R)^3
\end{equation} 
with $L=8$. Here we have already eliminated the non--Abelian symmetries by
forming \SU{N} gauge invariant monomials (cf.\ our discussion in
\ref{sec:NonAbelian}). The discrete symmetries follow from the space--group
selection rules and $H$--momentum conservation \cite{Dixon:1986qv} (for the
rules in this specific geometry see \cite{Blaszczyk:2009in}), and are partially
redundant, e.g.\ there are in fact only two independent $\Z4^R$ factors.  These
symmetries and selection rules constrain the holomorphic correlators of the
theory \cite{Dixon:1986qv, Hamidi:1986vh},  which can also partially come from
the K\"ahler potential in an appropriate description (cf.\ the discussion in
\cite[section~4]{Brummer:2010fr}). In what follows we will not distinguish
between such correlators and allowed superpotential terms.

In the vacuum discussed in \cite{Kappl:2010yu}, there is a
residual $\Z4^R$ symmetry, which forbids the superpotential at the perturbative 
level.
If we switch on an  additional field with $R$ charge 2 we will break the $\Z4^R$
and obtain a  non--zero superpotential in the vacuum. We will switch on the
additional field  $\bar{\phi}_1$. 
That is, the fields
\begin{equation}
\begin{split}
\tilde{\phi}^{(i)}~=~\{&
\bar{\phi}_1,\phi_1,\phi_2,\phi_3,\phi_4,\phi_5,
\phi_6,\phi_7,\phi_8,\phi_9,\phi_{10},\phi_{11},\phi_{12},\phi_{13},\phi_{14},\\
&
x_1,x_2,x_3,x_4,x_5,\bar{x}_1,\bar{x}_3,\bar{x}_4,\bar{x}_5,y_3,y_4,y_5,y_6\}
\end{split}
\end{equation} 
will now acquire vacuum expectation values (VEVs).
The Hilbert basis contains 15408 elements for this choice. 
The superpotential starts at lowest order with 4 Hilbert basis elements
\begin{equation}
\mathscr{W}~=~(x_4\bar{x}_4+x_5\bar{x}_5+\phi_9\phi_{13}+\phi_{10}\phi_{14})\,
\bar{\phi}_1+\cdots\;.
\end{equation}
We also considered the appearance of the proton decay operator $QQQ\ell$. The 
Hilbert basis involving $QQQ\ell$ consists of 4284 elements where we focus only 
on first generation quarks and leptons. The lowest order $QQQ\ell$ operator occurs 
at order 11 in the field VEVs. An example is given by
\begin{equation}
\mathscr{W}~\supset~ Q_1\,Q_2\,Q_2\,\ell_1\, \bar{\phi}_1\,x_1\,x_2\,x_3\,x_4\,\bar{x}_3\,
\phi_2\,\phi_4\,\phi_9\,\phi_{12}^2\;.
\end{equation}
These examples show that the Hilbert basis method is powerful enough to handle 
very complex examples with many fields.

\section{Applications and speculations}
\label{sec:Applications}

Defining the subsets
$\mathscr{H}_\mathrm{inhom}=\{\mathscr{M}_\mathrm{inhom}^{(1)},\dots 
,\mathscr{M}_\mathrm{inhom}^{(H_1)}\}$ and
$\mathscr{H}_\mathrm{hom}=\{\mathscr{M}_\mathrm{hom}^{(1)},\dots 
,\mathscr{M}_\mathrm{hom}^{(H_0)}\}$ 
which are constructed from $n_\mathrm{inhom}^{(i)}$ and
$n_\mathrm{hom}^{(h)}$, the full superpotential 
to all orders is 
\begin{equation}
 \mathscr{W}
 ~=~
 \sum\limits_{i=1}^{H_1}
 \sum\limits_{n_1,\dots,n_{H_0}}
 \lambda_{n_1\cdots n_{H_0}}^{(i)}
 \mathscr{M}_\mathrm{inhom}^{(i)}
 \left(\mathscr{M}_\mathrm{hom}^{(1)}\right)^{\eta_1}\cdots 
 \left(\mathscr{M}_\mathrm{hom}^{(H_0)}\right)^{\eta_{H_0}}
 \;.
\end{equation}
We can further speculate that the structure of the superpotential  is
\begin{equation}\label{eq:Wstructure}
 \mathscr{W}_\mathrm{structure}~=~
 \frac{\sum\limits_{i=1}^{H_1}\lambda_i\,\mathscr{M}_\mathrm{inhom}^{(i)}}{%
 1-\sum\limits_{h=1}^{H_0}\kappa_h\,\mathscr{M}_\mathrm{hom}^{(h)}
 }\;.
\end{equation}
However, the relations between the couplings implied by
$\mathscr{W}_\mathrm{structure}$ will in general be incorrect. Yet
$\mathscr{W}_\mathrm{structure}$ can be used quickly to answer certain questions
such as at which order some combination of fields appears first. Such questions
arise in (generalized) Froggatt--Nielsen model building \cite{Froggatt:1978nt}
with many symmetries and where the order in which a given term appears is a
measure for the suppression of the corresponding effective coupling. Another,
very similar application concerns the identification of approximate continuous
$R$ and non--$R$ symmetries, where one is interested in the lowest order terms
that explicitly break such symmetries \cite{Kappl:2008ie,Choi:2009jt}. Given
\eqref{eq:Wstructure} one immediately reads off the lowest order at which
perturbative superpotential terms arise.

Whether or not the relations between the coefficients implied by
\eqref{eq:Wstructure} can be obtained in some interesting setting needs still to
be explored. However, it is tempting to speculate that in certain  highly
symmetric string compactifications, such as orbifolds, the superpotential (and
holomorphic terms in the K\"ahler potential) will be some known function of the
building blocks. It will be interesting to study this question in more detail.

\section{Conclusions}
\label{sec:Conclusions}

We have described a simple method that allows us, given the symmetries of the
theory, to construct the building blocks of the $D$--flat directions and the
superpotential. This basis is given by the basis of non--negative integer
solutions $n$ of the simple matrix equation $C\cdot n=0$, which has been
extensively studied in the mathematical literature. Publicly available codes
allow us then to  compute the basis very efficiently. We have discussed a
specific example, based on a $\Z2\times\Z2$ orbifold model, in which we computed
the superpotential basis for certain singlet fields. The successful construction
of this basis demonstrates that our methods can be used to efficiently compute
the superpotential to all orders even in rather complex systems.

\subsection*{Acknowledgments}

We thank  Raymond Hemmecke and Jonas Schmidt for interesting discussions, and
Patrick Vaudrevange for valuable comments. This research was supported by the
DFG cluster of excellence Origin and Structure of the Universe and the
Graduiertenkolleg ``Particle Physics at the Energy Frontier of New Phenomena''
by Deutsche Forschungsgemeinschaft (DFG). We would like to thank the CERN theory
group, where some of this work has been carried out, for hospitality and
support.

\appendix

\section{Gauge invariant monomials for $\boldsymbol{\SU{N}}$}
\label{app:NonAbelianMonomials}

We review the construction of gauge invariant monomials for $\SU{N}$ with matter
fields $\phi_{i}$ and $\overline{\phi_{j}}$.

\subsection{Consecutive basis building}
\label{subsec:ConBasis}

Let us consider a general example of $L$ $\SU{N_{i}}$ 
gauge groups, $\SU{N_{1}}\times
\dots \times \SU{N_{L}}$. In order to construct the Hilbert basis
$\mathscr{H}$ we proceed as follows:  in a first step, we construct a basis
$\mathscr{H}_{1}$ of $\SU{N_{1}}$ singlets, consisting of elementary
$\SU{N_{1}}$ singlets and $\SU{N_{1}}$ invariant monomials. 
As is well known, the latter will be given by the `mesons' and `baryons', 
which in $\SU3$ would look like,
\begin{equation}\label{eq:baryonsmesons}
(\phi_{i} \overline{\phi}_{j}) \equiv \phi^{\alpha}_{i} \overline{\phi}_{j}^{\dot{\alpha}} \delta_{\alpha \dot{\alpha}}
\;,
\\
(\phi_{i} \phi_{j} \phi_{k}) \equiv \phi_{i}^{\alpha} \phi_{j}^{\beta}
\phi_{k}^{\gamma} \varepsilon_{\alpha \beta \gamma}\;,
\\
(\overline{\phi}_{i} \overline{\phi}_{j} \overline{\phi}_{k}) \equiv \overline{\phi}_{i}^{\dot{\alpha}} 
\overline{\phi}_{j}^{\dot{\beta}} \overline{\phi}_{k}^{\dot{\gamma}}
\varepsilon_{\dot{\alpha} \dot{\beta} \dot{\gamma}}\;.
\end{equation}
These monomials will transform as singlets under $\SU{N_{1}}$ and together with
the singlet fields they build the basis $\mathscr{H}_{1}$. 

We now use $\mathscr{H}_{1}$ to build the basis $\mathscr{H}_{1,2}$, which will
be the basis of $\SU{N_{1}}\times\SU{N_{2}}$ singlets. Obviously,
$\mathscr{H}_{1,2}$ will contain terms of $\mathscr{H}_{1}$ which are already
$\SU{N_{2}}$ singlets and also monomials which are constructed in a similar way
as in \eqref{eq:baryonsmesons}. The only difference is that terms in
$\mathscr{H}_{1}$ can have more than one $\SU{N_{2}}$ index, 
i.e.\ they can furnish higher representations.

From here on the course of action is always the same: we use the previous basis
$\mathscr{H}_{1,\dots,k}$ to build $\mathscr{H}_{1,\dots,k,k+1}$, the basis of
$\SU{N_{1}}\times \dots \times\SU{N_{k}}\times\SU{N_{k+1}}$ singlets
until we eventually find the basis $\mathscr{H} \equiv \mathscr{H}_{1,\dots,L}$
of monomials invariant under the full gauge group. One can find an explicit
example of this method for $\SU3\times\SU2\times\U1$ in
\cite{Gherghetta:1995dv}.

\subsection{Cartan subalgebras}

Alternatively, one can construct the monomials for non--Abelian symmetry groups
by using the Hilbert basis method for the $\U1^L$ symmetry defined by the Cartan
subalgebras. For this strategy, we split all fields in their tensor components
and assign them a charge according to the Cartan charges. Let us consider some
fields from \cite{Gherghetta:1995dv} to explain this procedure. Assume, we have
an $\SU3\times\SU2$ gauge group. \SU3 has rank 2 and its Cartan subalgebra can
be taken to be the one generated by the two diagonal Gell--Mann matrices,
$\lambda_{3}$ and  $\lambda_{8}$,
whereas \SU2 has rank 1, hence we use the diagonal Pauli matrix $\sigma_{3}$.

Let us assume we have three fields, $Q(\boldsymbol{3},\boldsymbol{2})$,
$\overline{u}(\boldsymbol{\overline{3}},\boldsymbol{1})$ and
$\ell(\boldsymbol{1},\boldsymbol{2})$. $Q$ has six tensor components $Q^{\alpha
i}$, where $\alpha=1,2,3$ is the \SU3 index and $i=1,2$ the \SU2 index. Now we
can assign charges to each component under the respective gauge groups, 
using the eigenvalues of the generators of the  Cartan subalgebras, which is
particularly easy when using the diagonal  matrices  $\lambda_{3},\lambda_{8}$
and $\sigma_{3}$. Therefore, $Q_{2} \equiv Q^{\alpha=2, i=1}$ will be assigned
the charges
\begin{equation}
q_{1}^{Q_{2}}~ = ~-1\;, \quad
q_{2}^{Q_{2}}~ =~ 1\;, \quad 
q_{3}^{Q_{2}}~=~1\;.
\end{equation}

Another example is $\overline{u}^{\dot{\alpha}}$, which is an \SU2 singlet and
an \SU3 anti--triplet, thus carrying zero charge under the SU(2) gauge group
and opposite SU(3) charges. Taking the component $\overline{u}_{2} \equiv
\overline{u}^{\dot{\alpha}=\dot{2}}$, we get 
\begin{equation}
q_{1}^{\overline{u}_{2}} 
=1\;, \\ q_{2}^{\overline{u}_{2}} 
=-1\;, \\ q_{3}^{\overline{u}_{2}}=0
\;.
\end{equation}
In this way we can split each field into its components (six for $Q$, three for
$\overline{u}$ and two for $\ell$) and build a $3 \times 11$ charge matrix  $Q$, which in our
example will look like
\begin{equation}
Q ~=~ \left(\begin{array}{ccccccccccc}
  1 & -1 & 0 & 1 & -1 & 0 & -1 & 1 & 0 & 0 & 0 \\
  1 & 1 & -2 & 1 & 1 & -2 & -1 & -1 & 2 & 0 & 0\\
  1 & 1 & 1  & -1& -1 & -1 & 0 & 0 & 0 & 1 & -1
 \end{array}\right)\;.
\end{equation}
Using algorithms like \cite{4ti2, Normaliz}, we can now find the
Hilbert basis of all solutions for the charge matrix $Q$, the same way as before in
equation~\eqref{eq:MatrixQ}. 
Solutions would for example be $Q^{11}\, Q^{2 1}\, Q^{3 2}\, \ell^{2}$,
$\overline{u}^{\dot{1}}\overline{u}^{\dot{2}}\overline{u}^{\dot{3}}$, $Q^{1 1} \overline{u}^{\dot{1}} \ell^{2}$ or many more. 

In order to associate these solutions to proper gauge invariant monomials, we define a prescription of how to translate such 
expressions to objects in which the (generalized) color indices are contracted appropriately. An increasing series will
be contracted with the total antisymmetric Levi--Civita tensor, indices which
have the same value but are dotted and undotted will be contracted with the
Kronecker delta. Using this procedure, our examples would look like
\begin{subequations}
\begin{equation}
 Q^{1 1}\, Q^{2 1}\, Q^{3 2}\, \ell^{2} 
 \quad\Longleftrightarrow\quad 
 Q^{\alpha a}\, Q^{\beta b}\, Q^{\gamma c}\, \ell^{d}\, 
 \varepsilon_{\alpha \beta \gamma}\, \varepsilon_{ac}\, \varepsilon_{bd}\;,
\end{equation}
where we had to contract antisymmetrically several times, 
\begin{equation} \label{eq:UUUmonomial}
 \overline{u}^{\dot{1}}\,\overline{u}^{\dot{2}}\,\overline{u}^{\dot{3}}
 \quad \Longleftrightarrow\quad 
 \overline{u}^{\dot{\alpha}}\, \overline{u}^{\dot{\beta}}\,\overline{u}^{\dot{\gamma}}\,
 \varepsilon_{\dot{\alpha}\dot{\beta}\dot{\gamma}}\;,
\end{equation}
which is very similar to the one above and
\begin{equation}
 Q^{1 1}\, \overline{u}^{\dot{1}}\, \ell^{2} 
 \quad\Longleftrightarrow\quad 
 Q^{\alpha a}\, \overline{u}^{\dot{\alpha}}\, \ell^{b}\, 
 \delta_{\alpha \dot{\alpha}}\, \varepsilon_{ab}\;,
\end{equation}
\end{subequations}
where we were able to see when to use the Kronecker delta.

A caveat of this procedure is that we will end up with many monomials occurring
more than once, e.g.\ \eqref{eq:UUUmonomial} will appear six
times. Therefore, we have to remove the redundant ones. Furthermore, one
has the possibility to end up with monomials which will vanish due to
antisymmetric contraction. Take \eqref{eq:UUUmonomial} again: if $\overline{u}$ would be a
field with only one generation, the monomial would clearly vanish. This means,
one has to check the Hilbert basis for zero--valued monomials and remove them,
which will leave the basis nonetheless intact, since the contribution of these
monomials would be zero in any case. 

We see that all these caveats are manageable. Furthermore, this procedure has a
big advantage compared to the method described in the previous subsection
\ref{subsec:ConBasis}. Using the Cartan subalgebras allows us to quickly and
fully automated build monomials for several \SU{N} and \U{1} gauge groups, which
would get very tedious (in certain cases impossible), especially for $N>3$, more
than three gauge groups or too many fields. We have created a
\texttt{mathematica} package ourselves, based on \cite{4ti2, Normaliz}, which
automatizes this procedure \cite{Dcode:2011df}.

\bibliography{Orbifold}
\bibliographystyle{NewArXiv}

\end{document}